# The DURATIONS randomised trial design: estimation targets, analysis methods and operating characteristics

**Running head: DURATIONS** design inference

**Word count:** 3906

**Authors:** Matteo Quartagno, James R. Carpenter, A. Sarah Walker, Michelle Clements, Mahesh K.B. Parmar

**Affiliations:** Institute for Clinical Trials and Methodology, University College London, 90 High Holborn, WC1V 6LJ

**Corresponding author:** Matteo Quartagno, MRC Clinical Trials Unit, University College London, 90 High Holborn, Second Floor. WC1V 6LJ, London, UK.
Email: m.quartagno@ucl.ac.uk

**Abstract:**

**Background.** Designing trials to reduce treatment duration is important in several therapeutic areas, including TB and antibiotics. We recently proposed a new randomised trial design to overcome some of the limitations of standard two-arm non-inferiority trials. This DURATIONS design involves randomising patients to a number of duration arms, and modelling the so-called 'duration-response curve'. This article investigates the operating characteristics (type-1 and type-2 errors) of different statistical methods of drawing inference from the estimated curve.

**Methods**. Our first estimation target is the shortest duration non-inferior to the control (maximum) duration within a specific risk difference margin. We compare different methods of estimating this quantity, including using model confidence bands, the delta method and bootstrap. We then explore the generalisability of results to estimation targets which focus on absolute event rates, risk ratio and gradient of the curve.

**Results.** We show through simulations that, in most scenarios and for most of the estimation targets, using the bootstrap to estimate variability around the target duration leads to good results for DURATIONS design-appropriate quantities analogous to power and type-1 error. Using model confidence bands is not recommended, while the delta method leads to inflated type-1 error in some scenarios, particularly when the optimal duration is very close to one of the randomised durations.

**Conclusions.** Using the bootstrap to estimate the optimal duration in a DURATIONS design has good operating characteristics in a wide range of scenarios, and can be used with confidence by researchers wishing to design a DURATIONS trial to reduce treatment duration. Uncertainty around several different targets can be estimated with this bootstrap approach.

# 1. Introduction

In several therapeutic areas, it is important to identify the optimal duration of treatment, defined as the shortest duration providing an acceptable efficacy. For example, reducing antibiotic treatment duration has been suggested as a way of combatting antimicrobial resistance[1], but this has to be done while maintaining high cure rates. Furthermore, shorter treatment durations often increase adherence, reduce side-effects and will be more cost-effective, provided they do not lead to an increased risk of relapse.

We recently proposed the DURATIONS randomised trial design[2] as an improvement over standard non-inferiority trials[3] for identifying the shortest acceptable treatment duration[4]. Its main attraction is that it does not involve selection of a single shorter duration to test against a control, which is often chosen based on very limited prior evidence; instead, patients are randomised to multiple durations, enabling the relationship between duration and response to be directly estimated using pre-specified flexible regression models. Randomising 500 patients to 5-7 durations enables the underlying duration-response curve to be estimated within 5% average absolute error in 95% of simulations[2].

The DURATIONS design moves away from a binary hypothesis testing paradigm, so that the trial outcome is not a treatment difference between two fixed durations, but rather the whole estimated duration-response curve. In applications, we wish to use this curve to inform decision-making, and hence it is essential to understand the properties of various ways of drawing inference from the curve. This article therefore first defines DURATION-design quantities analogous to power and type-I error, and then compares different strategies of inference for these quantities, through extensive simulations.

## 2. Drawing inference from the duration-response curve

Suppose there is a treatment that is known to be highly effective compared to placebo, and that is usually prescribed as standard-of-care to patients with a particular disease or condition for a fixed course duration; examples include antibiotics for bacterial infections and direct acting antivirals against Hepatitis C. Additionally, suppose the recommended treatment course is $D_{MAX}$ days, although we believe that shorter durations might be similarly effective, so that we suspect the shortest effective duration might be as small as $D_{MIN}$.

Now, suppose we want to design a trial to identify the shortest effective duration and assume that the primary outcome of the trial is cure, a binary variable equal to 1 if the patient recovers from their condition or 0 otherwise. Using a DURATIONS randomised trial design[2] we randomise N, say 500, patients to multiple, e.g. 7, duration arms, between and including $D_{MAX}$ and $D_{MIN}$. After observing the responses, we fit a pre-specified fractional polynomial logistic model with cure as outcome and duration as the only covariate, with up to two polynomial terms. This gives us an estimated duration-response curve, similar to the black curve in Figure 1.

When provided with this curve, how should clinicians and policy-makers choose what is the 'optimal' duration to prescribe? A simple choice is to target the duration leading to at most a fixed risk loss of efficacy (difference $\delta$) compared to the control (maximum) duration, e.g. 5% less. Hence, if the estimated control cure rate at $D_{MAX}$ was $\pi_{D_{MAX}}$, our objective would be to find the minimum duration $D^*$ corresponding to a cure rate of at least $\pi_{D^*} = \pi_{D_{MAX}} - \delta$. The rationale for this choice closely corresponds to that for the choice of margin in non-inferiority trials, i.e. it answers the question "what is the minimum treatment efficacy that we would be happy to tolerate, given the ancillary advantages of the active treatment (i.e. of reducing treatment duration)"?

In this section we propose different ways of estimating such a $D^*$ from the duration-response curve, and investigate their operating characteristics in Section 3. We extend to different possible objectives in Section 4.

### *2.1. Model confidence bands*

The simplest approach is to extrapolate $D^*$ from the duration-response curve, selecting the duration at which $\pi_{D^*} = \pi_{D_{MAX}} - \delta$. However, the fitted curve is just the most likely estimated from the sample data. Thus, there is a non-negligible probability that the true $D^*$ is below this value. Instead, we will typically wish to keep type-1 error below 2.5%, so that we do not recommend a duration that is not long enough more than 2.5% of the time.

The first and most naïve method of controlling type-1 error uses the lower bound of the pointwise confidence bands around the curve, looking for the duration $D^*$ satisfying $\pi_{D^*} - 1.96 SE_{\pi_{D^*}} = \pi_{D_{MAX}} - \delta$. To avoid recommending non-integer durations, which might make little sense in applications, we round the selected duration $D_L^*$ up to the next whole number; hence, we select the duration $D_I$ that satisfies:

$$D_I = \min_{D_i}(D_i > D_L^*)$$

With $\boldsymbol{D_i}$ being the set of integer durations.

## 2.2. Delta method CI

To compare two specific points on the duration-response curve, we need to estimate a confidence interval around the difference in outcome between the two points. In our application, we really want to compare each shorter duration $D_i$ against the longest duration $D_{MAX}$; hence, we want to estimate the confidence intervals around $\pi_{D_{MAX}} - \pi_{D_i}$ for every $D_i$ (including any integer duration randomised to or not randomised to below $D_{MAX}$). Let us call these confidence intervals $CI^{1-\alpha}_{D_{MAX}-D_i} = \left(CI^l_{D_{MAX}-D_i}, CI^u_{D_{MAX}-D_i}\right)$, with $\alpha$ being the confidence level. These can be estimated using the delta-method, which gives a good approximation based on Taylor series expansions (details in Appendix A). Inference can then be drawn by selecting the minimum duration $D^*$ for which $CI^u_{D_{MAX}-D^*} < \delta$.

## 2.3. Bootstrap CI

Alternatively, rather than relying on these approximations, given the observed dataset with N observations we can sample with replacement N observations M times, generating M bootstrap samples. We then fit the fractional polynomial model on each of the M datasets and estimate in each bootstrap sample $\pi_{D_{MAX}} - \pi_{D_i}$ for every $D_i$. We then calculate bootstrap confidence intervals around these quantities, and use these to choose the optimal $D^*$.

## 2.4. Bootstrap Duration CI

Another way of using bootstrapping is to directly estimate a confidence interval around $D^*$. M bootstrap samples are selected similarly to Section 2.3, but instead of estimating $\pi_{D_{MAX}} - \pi_{D_i}$ in each sample we estimate the corresponding $D^*$, denoted $D^*_m$, with $m$ indexing the bootstrap sample. Then a bootstrap confidence interval $(D^{*,l}, D^{*,u})$ can be constructed from the bootstrap mean estimate $\widehat{D^*}$ and its standard error, and the recommended duration would be:

$$D_I = \min_i( D_i > D^{*,u})$$

## 2.5. Theoretical comparison of methods

Table 1 provides an overview of the properties of different methods. The attractiveness of the confidence bands method comes from its simplicity, as it is probably the method most researchers would naturally use to estimate $D^*$ from the duration-response curve. However, it has several limitations, the most important being that the pointwise confidence bands for the curve are not the same as the confidence interval for the difference between two specific points on the curve.

While the delta method (2.2) addresses this problem, it is affected by at least one other issue, at least when using fractional polynomials[5] as the flexible regression method: specifically, that model selection variability should be taken into account[6]. The two bootstrap methods in Sections 2.3 and 2.4 are theoretically appealing strategies to solve this problem, as repeating the fractional polynomial selection step for each bootstrap sample is one approach to address model selection variability.

However, with the delta method (2.2) and the bootstrap CI method (2.3) we estimate multiple confidence intervals around the curve, and compare each upper bound against the maximum tolerable risk difference. Hence, we may theoretically run into a multiplicity issue. However, our repeated tests are performed with the goal to solve an equation, rather than to formally compare different duration arms, making this less problematic.

When the model used to estimate the duration-response curve is correct, we expect the bootstrap duration CI method (2.4) to estimate a confidence interval for $D^*$ that covers the true value at the nominal level. The assumption about model correctness is important; our choice of fractional polynomials as the preferred analysis method was originally driven by the fact that in many situations we are not confident what the true underlying model is, and hence flexible models are preferred. While the standard fractional polynomial algorithm was built as a parsimonious modelling technique[7], in our application, with a single covariate and a reasonable number of expected events, a modified algorithm selecting *exactly* two polynomial terms rather than a *maximum of* two is likely to be preferable, and makes it easier to satisfy the assumption that the model used is approximately correct. Henceforth, we refer to this method as the modified fractional polynomial analysis.

## 3. Simulations

We compared the methods presented above in a simulation study designed using the recently proposed ADEMP framework[8].

### *3.1. Aims*

The main aim is to compare different strategies of drawing inference from the duration-response curve. Specific questions are: are bootstrap methods necessary, or does the simpler delta-method suffice? Is the modified fractional polynomial analysis preferable? How problematic is the multiplicity issue with the bootstrap CI method?

### *3.2. Data-generating mechanisms*

The DURATIONS design aims to be resilient to the true underlying duration-response relationship. Consequently, we generate data under sixteen different scenarios that are listed and plotted in the additional material online, including the eight scenarios in (Quartagno et al, 2018). These reflect a wide range of possible duration-response relationships, including both those generated by fractional polynomials and those generated from sigmoid functions (which are not strictly within the fractional polynomial paradigm). For comparisons, we re-scale the x-axis so that the minimum duration considered ($D_{MIN}$) is 8 days, and the maximum $D_{MAX}$ is 20 days, with 7 randomised arms evenly spread between them (8, 10, 12, 14, 16, 18, 20). We generate 1000 data sets from each scenario of N=500 individuals. Using the same process, we additionally generate 200 datasets of N=750 and N=1000 individuals, to explore the sensitivity of results to total sample size.

### *3.3. Estimation targets*

We assume the estimation target of interest is the duration $D^*$ leading to a cure rate that is $\delta\%$ less than at the longest (and currently used) duration $D_{MAX}$. This approximately corresponds to the shortest duration non-inferior to standard-of-care within a non-inferiority margin $\delta$ on the risk difference scale. In our base case, we consider $\delta = 10\%$, a margin that has been recommended for example by FDA for non-inferiority trials of antibiotics for adult bacterial community acquired pneumonia[9]. However, as in some applications this may be considered too large a margin, we additionally explore the sensitivity of results using $\delta = 5\%$. This risk difference based estimation target might not be optimal in all applications, and hence in Section 4 we propose different estimation targets.

### *3.4. Methods*

We compare the four different methods of estimating $\widehat{D^*}$ described above. For the bootstrap methods, we use M=500 replications. For method 2.3, we build confidence intervals using the BCa method[10], which is based on the percentiles of the bootstrap distribution, but with adjustments to account for both bias and skewness. For method 2.4, we use the percentile confidence interval instead, as BCa is not applicable in some scenarios where the bootstrap distribution is truncated at 8 days. Additionally, we compare the standard (i.e. selecting up to two polynomial terms) and modified (i.e. selecting exactly two terms) fractional polynomial algorithms for method 2.4.

### 3.5. Performance measures

The choice of performance measures is not straightforward. Unlike most randomised trial designs, our DURATIONS randomised design does not define a null hypothesis to test. This is because the longest treatment duration we randomise patients to is already known to be (usually highly) effective, and our aim is only to find the most appropriate shorter duration. Under very simple assumptions, i.e. that the duration-response curve is continuous and monotonic, we know that either there is a duration $D^*$ in the interval $[D_{MIN}; D_{MAX}]$ such that $\pi_{D^*} = \pi_{D_{MAX}} - \delta$, or that even the shortest duration is sufficiently effective, i.e. $\pi_{D^*} = \pi_{D_{MIN}} \geq \pi_{D_{MAX}} - \delta$.

Given this, and having defined our estimation target, we define three main performance measures of interest:

1) Type-1 Error: in this context we define type-1 error as the probability that our trial ends up recommending a duration $\widehat{D^*}$ that is insufficiently effective:

$$Type\ 1\ Error = P(\pi_{\widehat{D^*}} < \pi_{D_{MAX}} - \delta)$$

2) Full Power: there are several ways to define power compared to standard hypothesis testing. We define *full power* as the probability that the trial ends up recommending the actual optimal duration, that is the minimum effective (integer) duration:

$$Full\ Power = P(\widehat{D^*} = \min_{D_i}(\pi_{D_i} \geq \pi_{D_{MAX}} - \delta))$$

3) Partial Power: designing a trial to reach high levels of Full Power might be difficult, but one might be interested in simply finding an effective duration that is shorter than the recommended one, even if it is not necessarily the *minimum* effective duration. We define the *partial power* as the probability that our trial identifies one such duration:

$$Partial\ Power = P(\pi_{\widehat{D^*}} \geq \pi_{D_{MAX}} - \delta)$$

Additionally, we measure performance by considering the distribution of all the recommended durations $\widehat{D^*}$.

## 3.6. Results

Figure 2 compares type-1 error and partial power across the different methods. It is immediately clear that type-1 error is not adequately controlled under certain scenarios. These scenarios are mainly those where the curvature at the optimal duration is positive, i.e. those for which steepness of the curve is increasing at the optimal duration. These are arguably not the most likely scenarios in our settings, as we expect to be investigating part of the duration-response curve where the curve is asymptoting; nevertheless, it is preferable to use methods like those in Section 2.3 and 2.4 that provide better inference by taking into account model selection variability. Differences between the two bootstrap methods are less marked (Table 2), although type-1 error is generally slightly lower using the bootstrap duration CI method (2.4).

In terms of analysis model, the modified fractional polynomial method is preferable with regards to type-1 error (Figure 2 and Table E in the additional material). There are only a few scenarios under which Methods 2.3 and 2.4 fail to control the type-1 error within 2.5%, namely scenarios 7, 8, 13 and 14 (Figure 3). However, this figure shows that the difference between the estimated minimum duration and the actual minimum acceptable duration is small (<=1 day) in these scenarios and that the 2.5$^{th}$ percentiles of recommended durations (and associated cure rate) are actually very close to the optimal duration (and corresponding cure rate).

In terms of power, differences are not as pronounced, and all methods achieve very good partial power (>90%) under most scenarios. Of note, the simulated sample size (N=500) was determined to estimate the duration-response curve within a certain average absolute error (5%), and not to achieve a specific power under any of our scenarios. As expected, full power is always substantially lower, particularly in scenarios where the cure rate at the optimal duration is very close to the minimum acceptable cure rate. Again, differences between different methods are not substantial.

The only scenario where full power exceeds 80% is Scenario 4 (Table 2, Tables A-D in additional material), i.e. constant cure rate at every duration. In this scenario, full power can be broadly interpreted as the power of a non-inferiority trial of $D_{MAX}$ against $D_{MIN}$ with a $\delta$ % non-inferiority margin, where the two durations have exactly the same true cure rate.

Finally, conclusions are similar using $\delta = 5\%$ (Figure a in additional material), although power is lower in all scenarios. In particular, partial power remains quite low for Scenario 14, only crossing the 80% threshold for N=1000. Full power remains very low for all scenarios and sample sizes.

## 4. Alternative objectives

In this simulation study, we assumed that the trial objective was to find the minimum duration $D^*$ such that $\pi_{D^*} \geq \pi_{D_{MAX}} - \delta$. This clearly corresponds with standard non-inferiority trials designed with margin $\delta_{NI}$ on the risk difference scale: in both, we assume that if the cure rate in the intervention arm is poorer only within a certain limit, then its secondary advantages make it preferable to the control arm. In this section we discuss different objectives. Some are again in common with standard non-inferiority trials, while others are specific to our design.

### 4.1. Fixed rate

Instead of considering a fixed difference in cure rate compared to control, one might simply want to identify the duration $D^*$ leading to a specific cure rate $\pi^*$:

$$D^* = \min_{D_i}(\pi_{D_i} \geq \pi^*)$$

This may be reasonable when there is good prior information on the expected cure rate $\pi_{D_{MAX}}$ at the control duration $D_{MAX}$, although one can never be sure that the population recruited will reflect historical controls. Methods 2.3 and 2.4 could be used as analysis methods for this estimation target.

### 4.2. Fixed risk ratio

Analogously to a standard non-inferiority trial, the margin $\delta$ could also be defined on a relative scale; for example, as the proportion of the cure rate at $D_{MAX}$ that should be preserved in order to consider the shorter duration preferable:

$$\min_{D_i}(\pi_{D_i} \geq \delta * \pi_{D_{MAX}})$$

For example, if we wanted to preserve 90% of the effect of treatment at $D_{MAX}$, we would choose $\delta = 0.9$.

### 4.3. The acceptability frontier

While non-inferiority trials naturally have a single non-inferiority margin $\delta_{NI}$, in DURATIONS trials it might instead be logical to have different margins for each specific duration $D_i$. For example, for a duration that is a half of $D_{MAX}$, we might be happy to tolerate a slightly larger $\delta$ than for a duration equal to two thirds of $D_{MAX}$, assuming that advantages of shorter durations generally increase as the duration reduces. Hence, the objective of the trial might be to find the duration $D^*$ such that:

$$D^* = \min_{D_i}(\pi_{D_i} \geq \pi_{D_{MAX}} - \delta(D_i))$$

where $\delta(D_i)$ is the function that indicates the acceptable loss in cure rate for each duration $D_i$. We call this the *Acceptability Frontier,* a possible example of which is the grey line in Figure 1.

### 4.4. Maximum gradient

Instead of targeting a specific cure rate, or difference in cure rate, investigators might be interested in the gradient of the duration-response curve. If we expect the curve to asymptote after a certain duration D*, then we might define such D* as the point after which the gradient of the function is always below a certain threshold $\delta$:

$$D^* = \min_{D_i}(\nabla \pi(D) \leq \delta, \forall D > D_i)$$

For example, in the scenarios above, we found that $\delta = 2\%$ was a reasonable value for this threshold.

### *4.5. Additional simulations*

We performed additional simulations to explore operating characteristics of the bootstrap duration CI method (2.4) with each of these different estimation targets. We analysed the same 1000 datasets generated for the 16 scenarios in our base case simulations, with N=500. As the target cure rate, we chose in each scenario the value 10% less than the true cure rate at the longest duration. For the fixed risk ratio, we used $\delta = 0.9$. The acceptability frontier was a linear function, such that the largest acceptable risk difference at $D_{MIN}$ was 10%, and at 18 days 5% (Figure 1). Finally, the maximum acceptable gradient was taken to be 2%. For this last estimation target we did not use method 2.4, but we simply looked at the single point estimate in the data. Results ( Tables D-G in the additional material) show reasonable performance for the first three alternative estimation targets. For the maximum gradient target, type-1 errors are large, and further analysis refinements would be necessary to use this.

## 5. Analysis example

Given the simulation results, here we sketch our favoured three-step approach to analysing a DURATIONS trial. First, an acceptability frontier should be defined, answering the question "what would the non-inferiority margin be for a trial comparing the longest duration to each shorter one?". In this example, we assume a reasonable non-inferiority margin is 10% cure rate difference compared to the control duration, as in the base-case scenario of our simulations (Figure 4). Second, we run our modified fractional polynomial algorithm to estimate the duration-response curve (black solid line in Figure 4). Third, we use the bootstrap (BCa method[10]) to find the confidence interval either around our estimated optimal duration $\widehat{D^*}$ or the differences in cure rate from the control at each duration (left and right panel of Figure 4 respectively). In this example, both methods recommend $\widehat{D^*} = 13$ days as optimal duration.

## 6. Discussion

In this article we have compared different strategies for drawing inference from a duration-response curve estimated using a DURATIONS randomised trial design. We defined quantities analogous to type-1 error and power in this scenario, and found a method based on bootstrap re-sampling — to estimate the duration associated with a specific cure rate difference from control — has good inferential properties when combined with a modified fractional polynomial analysis method. This is therefore our recommended approach.

One issue with the standard non-inferiority design for identifying optimal treatment duration is the potential for so-called 'bio-creep', i.e. the erosion of efficacy from sequentially testing for non-inferiority shorter durations, iteratively updating the control duration to one previously shown to be non-inferior[11]. One advantage of the DURATIONS design is that it avoids this problem, as all the durations are evaluated in the same trial. Another advantage is its resilience; in a standard non-inferiority trial, whenever a single design parameter turns out to have been badly misjudged, the whole trial can quickly lose power or interpretability. By contrast, the DURATIONS design has been developed to be flexible enough to maintain good properties against a wide range of duration-response curves.

## 6.1. Extensions

Design and analysis of randomised trials often intersect, and hence what is an analysis decision (how to analyse and draw inference from the observed data) also has design implications (how to best design a trial that we aim to analyse in a particular way). Hence, future work will investigate how to design a DURATIONS randomised trial that we aim to analyse with methods presented here.

We focused on binary outcomes, but similar methods could be easily used for continuous and survival outcomes, with the only additional complication of having to derive a suitable estimation target. When developing the target based on the acceptability frontier, we assumed that this frontier was subjectively drawn by the analyst, based on assumptions about the trade-offs between shortening duration and loss of treatment efficacy. An alternative could be to derive the acceptability frontier using available data on the secondary advantages of shorter durations. For example, if the goal was to design a DURATIONS trial in Hepatitis C, the acceptability frontier could be built using data on costs.

Future work could include investigating the effect of including additional covariates in the fractional polynomial model, for example age or sex, if there was evidence that the optimal duration might vary depending on these factors. It could also investigate using an adaptive design, to allow for closure of poorly performing arms (i.e. those at the lowest durations)[4].

For TB and related settings, where the optimal duration might be investigated for a new drug or new regimen, it is important to investigate the best way to include a formal comparison with an independent control treatment of fixed duration, for example with the standard 6-month TB treatment course with rifampicin, isoniazid, pyrazinamide and ethambutol.

*6.2. Conclusions*

We recently proposed a DURATIONS randomised trial design as an alternative to a standard two-arm non-inferiority design when the goal is to optimise treatment duration. Here we have investigated the operating characteristics of various methods of drawing inference from the duration-response curve and found that a method based on bootstrap to estimate a duration associated with a specific cure rate has good properties and is an appealing choice. Using this analysis method, DURATIONS randomised trials could help identify better treatment durations in an optimal way across many illnesses[4].

**Code**

The code used for the simulations will be made available on the GitHub page of the first author.

**Funding**

This work was supported by the Medical Research Council [MC_UU_12023/29]. ASW is an NIHR Senior Investigator. The views expressed are those of the author(s) and not necessarily those of the NHS, the NIHR, or the Department of Health.

**ORCID iD**

Matteo Quartagno https://orcid.org/0000-0003-4446-0730

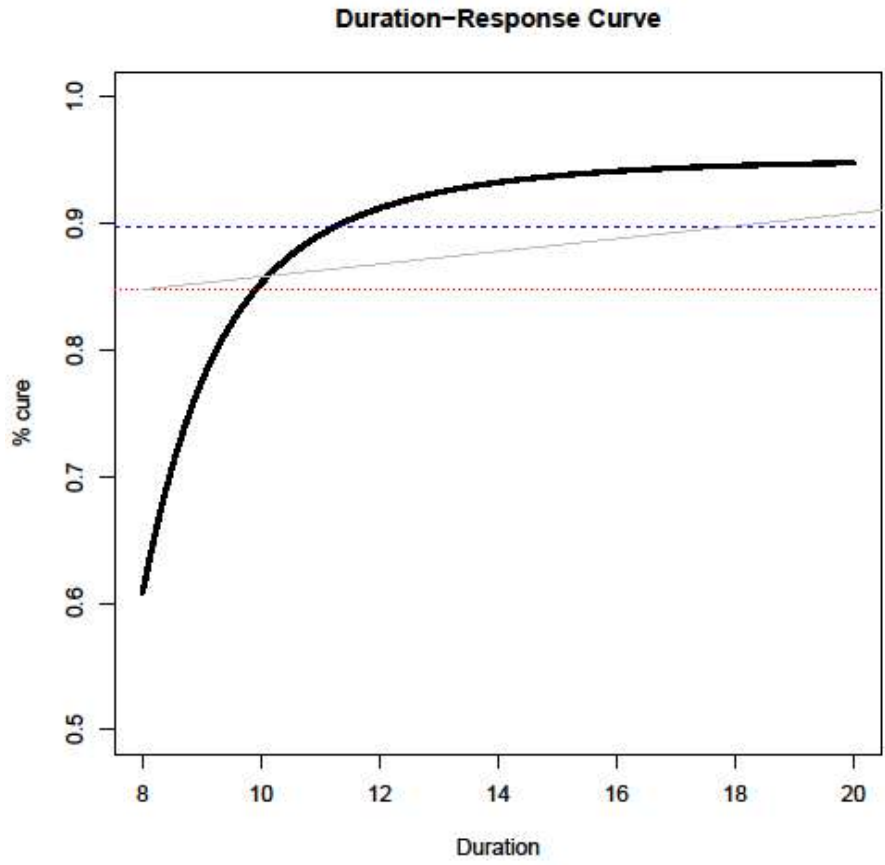

**Figure 1: Example of estimated duration-response curve (solid, black), drawn against three possible acceptability frontiers (dotted, red = 10% fixed risk difference; dashed, blue = 5% fixed risk difference; solid, grey= duration-specific frontier, as in section 4.3).**

Table 1: Properties of different methods to estimate confidence intervals.

| Method | Targets difference in efficacy between two points | Addresses model selection uncertainty | Addresses multiplicity issue | Analysis method used in simulations |
|---|---|---|---|---|
| Confidence Bands (§2.1) | No | No | No | Modified fractional polynomials (gamlss package in R) |
| Delta Method (§2.2) | Yes | No | No | Modified fractional polynomials (gamlss) |
| Bootstrap CI (§2.3) | Yes | Yes | No | Modified fractional polynomials (gamlss) |
| Bootstrap Duration CI (§2.4) | Yes | Yes | Yes | Both standard (mfp package) and modified (gamlss) fractional polynomials |

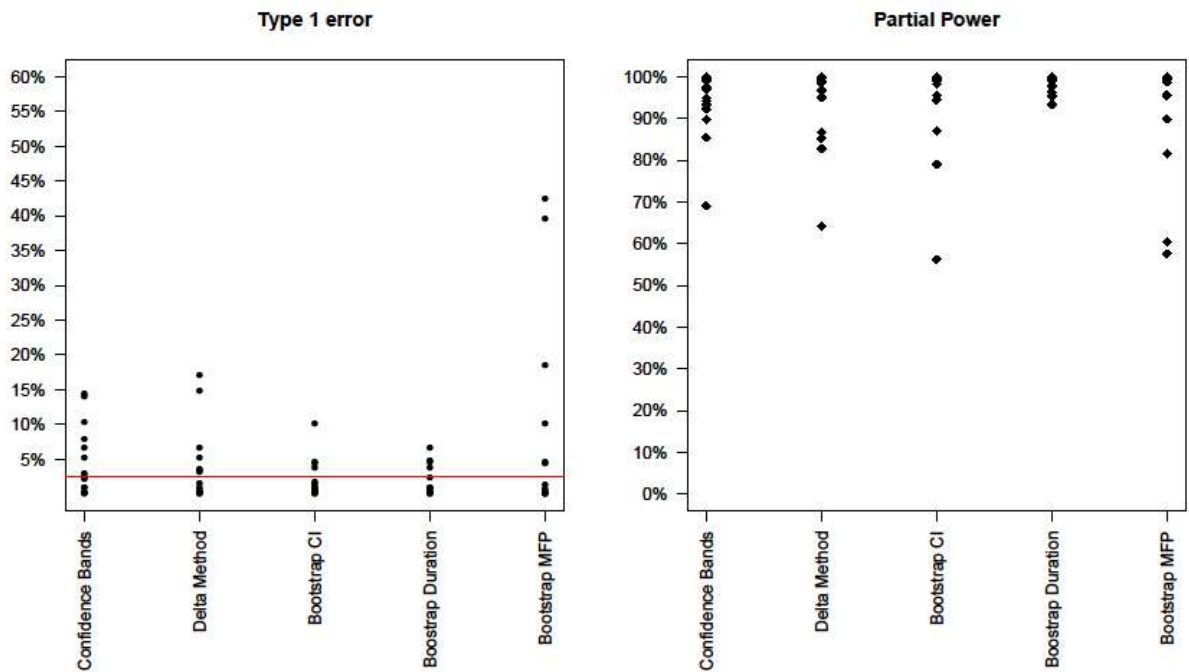

**Figure 2: Type-1 Error and Partial Power (probability of recommending any sufficiently effective shorter duration) of the 5 analysis methods across the 16 simulation scenarios. Bootstrap MFP uses the Bootstrap duration CI method, but using the standard fractional polynomial approach as the analysis method (as in R package mfp).**

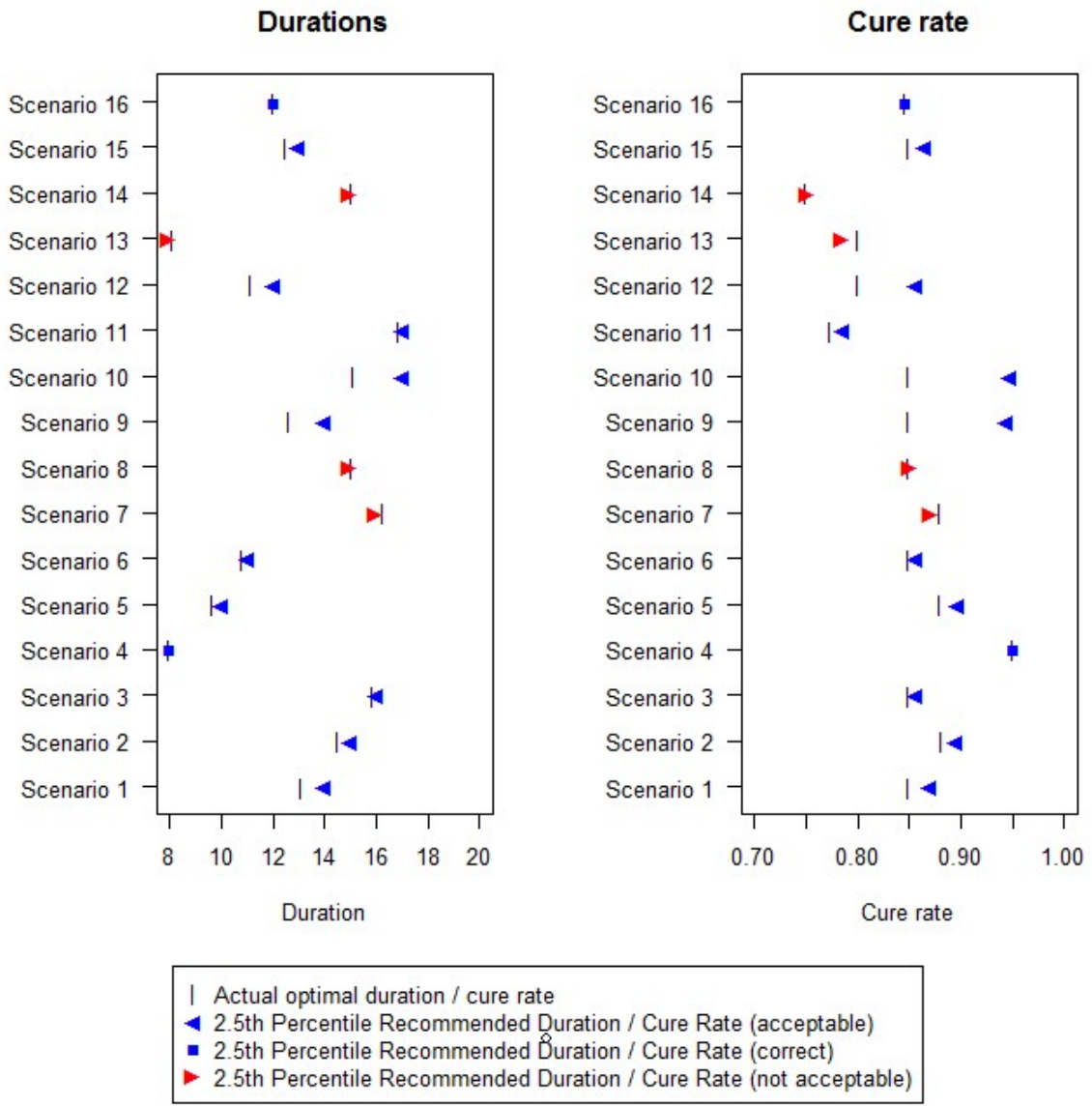

Figure 3: Duration recommended and associated cure rate across the 16 simulation scenarios using method 2.4, with base-case design parameters (estimation target=shortest duration non-inferior to 20 days within 10% risk difference). The vertical bar indicates the true minimum effective duration.

Table 2: Results from using methods 2.3 and 2.4, with the modified fractional polynomial analysis method and base-case design parameters (estimation target=shortest duration non-inferior to 20 days within 10% risk difference). In italics: scenarios for which the upper bound of the Monte-Carlo confidence interval for type-1 error was not strictly controlled within 2.5%, with Monte-Carlo confidence interval. See additional material Tables A-B for other methods.

|  | Partial Power (%) | Full Power (%) | Type-1 Error (%) (Target value: 2.5%) | True Min Duration | Estimated Minimum Duration Recomm. | Estimated 2.5th Perc Duration Recomm. | Estimated Median Duration Recomm. |
|---|---|---|---|---|---|---|---|
| **Method 2.3 (Bootstrap CI):** | | | | | | | |
| Scenario 1 | 98.2 | 9.8 | 1.8 | 13.1 | 12 | 14 | 16 |
| Scenario 2 | 99.2 | 16.5 | 0.8 | 14.5 | 14 | 15 | 16 |
| Scenario 3 | 94.5 | 10 | 1.6 | 15.9 | 15 | 16 | 18 |
| Scenario 4 | 100 | 82.7 | 0 | 8.0 | 8 | 8 | 8 |
| Scenario 5 | 99.6 | 5.4 | 0.4 | 9.7 | 9 | 10 | 12 |
| Scenario 6 | 99.7 | 3.2 | 0.3 | 10.8 | 10 | 11 | 14 |
| *Scenario 7* | *94.4* | *30.7* | *4.3 (3.0,5.6)* | *16.2* | *14* | *16* | *18* |
| *Scenario 8* | *87* | *12.9* | *10.1(8.2,12)* | *15.0* | *8* | *15* | *17* |
| Scenario 9 | 100 | 2.6 | 0 | 12.6 | 13 | 13 | 14 |
| Scenario 10 | 99.9 | 0.8 | 0 | 15.2 | 16 | 17 | 18 |
| Scenario 11 | 79 | 8.6 | 0.1 | 16.8 | 16 | 17 | 18 |
| Scenario 12 | 98.9 | 29.9 | 1.1 | 11.2 | 11 | 12 | 13 |
| *Scenario 13* | *95.5* | *26.6* | *4.5 (3.2,5.8)* | *8.1* | *8* | *8* | *10* |
| *Scenario 14* | *56.2* | *8.6* | *3.8 (2.6,5.0)* | *15.0* | *8* | *15* | *17* |
| Scenario 15 | 99.3 | 4.8 | 0.7 | 12.5 | 12 | 13 | 15 |
| Scenario 16 | 99.8 | 4.9 | 0.2 | 12.0 | 11 | 12 | 14 |
| **Method 2.4 (Bootstrap Duration CI):** | | | | | | | |
| Scenario 1 | 97.7 | 9.9 | 2.3 | 13.1 | 11 | 14 | 16 |
| Scenario 2 | 99.7 | 14.4 | 0.3 | 14.5 | 14 | 15 | 16 |
| Scenario 3 | 99 | 5.3 | 1 | 15.9 | 14 | 16 | 18 |
| Scenario 4 | 100 | 86.1 | 0 | 8.0 | 8 | 8 | 8 |
| Scenario 5 | 99.9 | 5.4 | 0.1 | 9.7 | 9 | 10 | 12 |
| Scenario 6 | 99.7 | 3.5 | 0.3 | 10.8 | 10 | 11 | 14 |
| *Scenario 7* | *95.2* | *46.8* | *4.8 (3.5,6.1)* | *16.2* | *15* | *16* | *17* |
| *Scenario 8* | *93.3* | *15.7* | *6.7 (5.2,8.2)* | *15.0* | *8* | *15* | *17* |
| Scenario 9 | 100 | 2.3 | 0 | 12.6 | 13 | 14 | 14 |
| Scenario 10 | 100 | 0.6 | 0 | 15.2 | 16 | 17 | 17 |
| Scenario 11 | 99.9 | 9.8 | 0.1 | 16.8 | 16 | 17 | 18 |
| Scenario 12 | 99 | 40.3 | 1 | 11.2 | 11 | 12 | 13 |
| *Scenario 13* | *96.3* | *29* | *3.7 (2.5,4.9)* | *8.1* | *8* | *8* | *10* |
| *Scenario 14* | *95.5* | *5.7* | *4.5 (3.2,5.8)* | *15.0* | *8* | *15* | *18* |
| Scenario 15 | 99.3 | 5 | 0.7 | 12.5 | 12 | 13 | 15 |
| Scenario 16 | 99.8 | 4.6 | 0.2 | 12.0 | 11 | 12 | 14 |

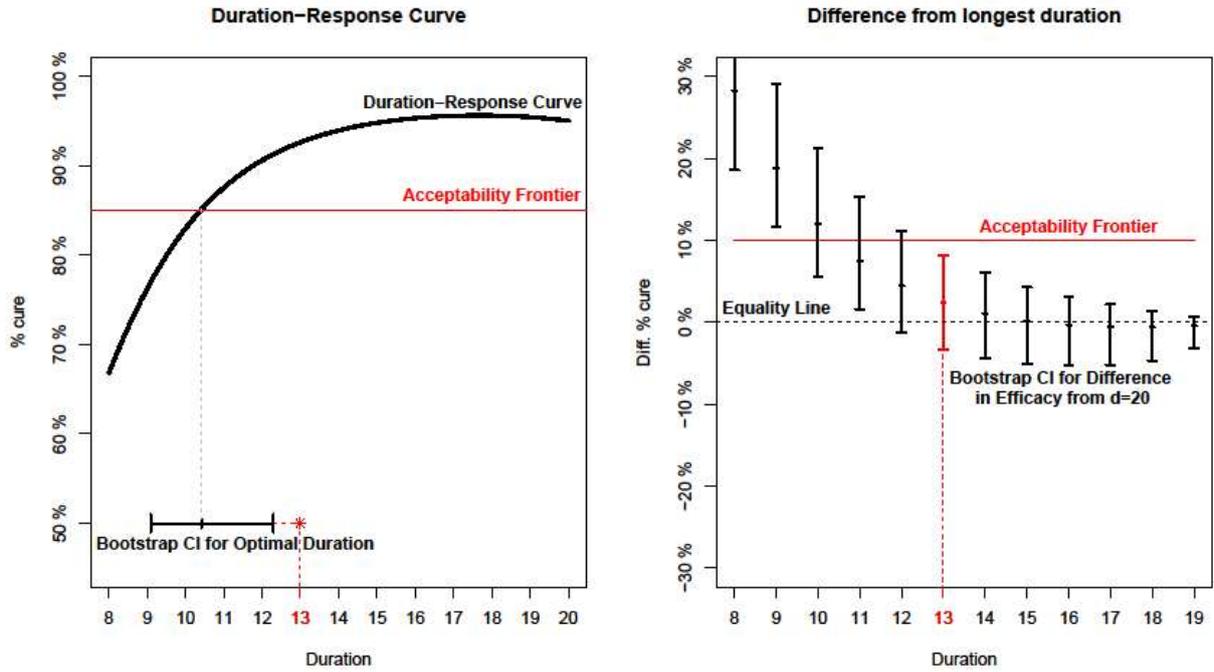

Figure 4: Analysis example for a hypothetical trial. On the left panel, the duration-response curve is estimated and then a bootstrap CI is built around the point where it crosses the acceptability frontier. On the right panel, bootstrap CIs are built around the difference in efficacy (cure rate) between each arm and the longest (d=20).

**Table A: Results from using method 2.1, Model Confidence Bands, with base-case design parameters (estimand=shortest duration non-inferior to 20 days within 10% risk difference)**

In italics: scenarios for which type 1 error was not strictly controlled within 2.5%. Note the standard fractional polynomial algorithm was used here, as mfp is the only R package returning standard errors for the data.

|  | Partial Power (%) | Full Power (%) | Type 1 Error (%) | True Min Duration | Estimated Minimum Duration | Estimated 2.5th Perc Duration | Estimated Median Duration |
|---|---|---|---|---|---|---|---|
| *Scenario 1* | *89.7* | *11.7* | *10.3* | *13.1* | *10* | *12* | *16* |
| Scenario 2 | 99.2 | 15.2 | 0.8 | 14.5 | 13 | 15 | 16 |
| Scenario 3 | 97.4 | 10 | 2.5 | 15.9 | 12 | 16 | 18 |
| Scenario 4 | 100 | 79.8 | 0 | 8.0 | 8 | 8 | 8 |
| Scenario 5 | 99.7 | 9.3 | 0.3 | 9.7 | 9 | 10 | 12 |
| Scenario 6 | 97.6 | 8.4 | 2.4 | 10.8 | 9 | 11 | 13 |
| *Scenario 7* | *93.3* | *23.2* | *6.7* | *16.2* | *14* | *16* | *18* |
| *Scenario 8* | *85.4* | *21.3* | *14.4* | *15.0* | *8* | *13* | *17* |
| Scenario 9 | 100 | 4.8 | 0 | 12.6 | 13 | 13 | 14 |
| Scenario 10 | 100 | 2.5 | 0 | 15.2 | 16 | 17 | 18 |
| Scenario 11 | 94.2 | 13.9 | 2.2 | 16.8 | 16 | 17 | 18 |
| *Scenario 12* | *97* | *53.5* | *3* | *11.2* | *11* | *11* | *12* |
| *Scenario 13* | *94.8* | *46.7* | *5.2* | *8.1* | *8* | *8* | *9* |
| *Scenario 14* | *69* | *14.1* | *13.9* | *15.0* | *8* | *11* | *17* |
| *Scenario 15* | *92.2* | *11.7* | *7.8* | *12.5* | *11* | *12* | *15* |
| Scenario 16 | 99.1 | 14.7 | 0.9 | 12.0 | 11 | 12 | 14 |

**Table B: Results from using method 2.2, Delta method CI, with base-case design parameters (estimand=shortest duration non-inferior to 20 days within 10% risk difference). Partial and full power, Type-1 Error, Real minimum duration, minimum, 2.5% percentile and median duration recommended. In italics: scenarios for which type 1 error was not strictly controlled within 2.5%.**

|  | Partial Power (%) | Full Power (%) | Type 1 Error (%) | True Min Duration | Estimated Minimum Duration | Estimated 2.5th Perc Duration | Estimated Median Duration |
|---|---|---|---|---|---|---|---|
| *Scenario 1* | *96.6* | *13.1* | *3.4* | *13.1* | *11* | *13* | *15* |
| *Scenario 2* | *96.8* | *29.5* | *3.2* | *14.5* | *14* | *14* | *16* |
| *Scenario 3* | *95.1* | *18.9* | *3.5* | *15.9* | *13* | *15* | *17* |
| Scenario 4 | 100 | 88.9 | 0 | 8.0 | 8 | 8 | 8 |
| Scenario 5 | 99.8 | 7.6 | 0.2 | 9.7 | 9 | 10 | 12 |
| Scenario 6 | 99.5 | 4.8 | 0.5 | 10.8 | 10 | 11 | 14 |
| *Scenario 7* | *85.2* | *46* | *14.8* | *16.2* | *14* | *16* | *17* |
| *Scenario 8* | *82.7* | *24.1* | *17* | *15.0* | *8* | *14* | *17* |
| Scenario 9 | 100 | 1.7 | 0 | 12.6 | 13 | 14 | 14 |
| Scenario 10 | 100 | 0.9 | 0 | 15.2 | 16 | 17 | 17 |
| Scenario 11 | 86.6 | 8.2 | 0.1 | 16.8 | 16 | 17 | 18 |
| Scenario 12 | 99 | 33.5 | 1 | 11.2 | 11 | 12 | 13 |
| *Scenario 13* | *94.9* | *26.1* | *5.1* | *8.1* | *8* | *8* | *10* |
| *Scenario 14* | *64.1* | *16.2* | *6.7* | *15.0* | *8* | *14* | *17* |
| Scenario 15 | 98.5 | 6.1 | 1.5 | 12.5 | 12 | 13 | 15 |
| Scenario 16 | 99.8 | 3.9 | 0.2 | 12.0 | 11 | 12 | 14 |

Table C: Results from using method 2.4, Bootstrap duration CI, but using standard fractional polynomials (mfp package in R), with base-case design parameters (estimand=shortest duration non-inferior to 20 days within 10% risk difference). Partial and full power, Type-1 Error, Real minimum duration, minimum, 2.5% percentile and median duration recommended. In italics: scenarios for which type 1 error was not strictly controlled within 2.5%.

|  | Partial Power (%) | Full Power (%) | Type 1 Error (%) | True Min Duration | Estimated Minimum Duration | Estimated 2.5th Perc Duration | Estimated Median Duration |
|---|---|---|---|---|---|---|---|
| Scenario 1 | 98.7 | 32.4 | 1.3 | 13.1 | 12 | 14 | 15 |
| *Scenario 2* | *95.6* | *44* | *4.4* | *14.5* | *14* | *14* | *16* |
| *Scenario 3* | *89.8* | *25.1* | *10.2* | *15.9* | *13* | *15* | *17* |
| Scenario 4 | 100 | 86.3 | 0 | 8.0 | 8 | 8 | 8 |
| Scenario 5 | 100 | 0.4 | 0 | 9.7 | 10 | 11 | 12 |
| Scenario 6 | 100 | 0.1 | 0 | 10.8 | 11 | 13 | 14 |
| *Scenario 7* | *81.6* | *41.5* | *18.4* | *16.2* | *14* | *15* | *17* |
| *Scenario 8* | *57.6* | *13.1* | *42.4* | *15.0* | *8* | *12* | *16* |
| Scenario 9 | 100 | 1.6 | 0 | 12.6 | 13 | 14 | 15 |
| Scenario 10 | 100 | 0.2 | 0 | 15.2 | 16 | 17 | 17 |
| Scenario 11 | 100 | 0.2 | 0 | 16.8 | 17 | 18 | 18 |
| Scenario 12 | 99.4 | 23.8 | 0.6 | 11.2 | 11 | 12 | 14 |
| *Scenario 13* | *95.4* | *9.3* | *4.6* | *8.1* | *8* | *8* | *11* |
| *Scenario 14* | *60.4* | *16.6* | *39.6* | *15.0* | *8* | *12* | *16* |
| Scenario 15 | 99.8 | 4.3 | 0.2 | 12.5 | 12 | 13 | 14 |
| Scenario 16 | 100 | 0.5 | 0 | 12.0 | 12 | 13 | 15 |

Table D: Results from using method 2.4, Bootstrap duration CI, targeting a fixed cure rate estimand. Partial and full power, Type-1 Error, Real minimum duration, minimum, 2.5% percentile and median duration recommended. In italics: scenarios for which type 1 error was not strictly controlled within 2.5%.

|  | Partial Power (%) | Full Power (%) | Type 1 Error (%) | True Min Duration | Estimated Minimum Duration | Estimated 2.5th Perc Duration | Estimated Median Duration |
|---|---|---|---|---|---|---|---|
| *Scenario 1* | *96.3* | *13.7* | *3.2* | *13.1* | *12* | *13* | *16* |
| Scenario 2 | 99.4 | 13.1 | 0.5 | 14.5 | 14 | 15 | 16 |
| Scenario 3 | 94.7 | 5.7 | 0.4 | 15.9 | 15 | 16 | 18 |
| Scenario 4 | 100 | 87 | 0 | 8.0 | 8 | 8 | 8 |
| Scenario 5 | 99.9 | 4.3 | 0.1 | 9.7 | 9 | 10 | 12 |
| Scenario 6 | 99.9 | 2.4 | 0.1 | 10.8 | 10 | 12 | 13 |
| *Scenario 7* | *93.9* | *36.3* | *3.3* | *16.2* | *15* | *16* | *18* |
| *Scenario 8* | *88.2* | *23.3* | *9.3* | *15.0* | *8* | *14* | *17* |
| Scenario 9 | 100 | 1.1 | 0 | 12.6 | 13 | 14 | 14 |
| Scenario 10 | 100 | 6.1 | 0 | 15.2 | 16 | 16 | 17 |
| Scenario 11 | 78.9 | 9.4 | 0.2 | 16.8 | 16 | 17 | 18 |
| Scenario 12 | 99.9 | 48.3 | 0.1 | 11.2 | 11 | 12 | 13 |
| *Scenario 13* | *97.2* | *43.8* | *2.8* | *8.1* | *8* | *8* | *10* |
| *Scenario 14* | *76.7* | *11.6* | *7.4* | *15.0* | *8* | *14* | *18* |
| Scenario 15 | 98.5 | 10.7 | 1.4 | 12.5 | 11 | 13 | 15 |
| Scenario 16 | 100 | 5.8 | 0 | 12.0 | 12 | 12 | 13 |

Table E: Results from using method 2.4, Bootstrap duration CI, targeting a fixed risk ratio estimand. Partial and full power, Type-1 Error, Real minimum duration, minimum, 2.5% percentile and median duration recommended. In italics: scenarios for which type 1 error was not strictly controlled within 2.5%.

|  | Partial Power (%) | Full Power (%) | Type 1 Error (%) | True Min Duration | Estimated Minimum Duration | Estimated 2.5th Perc Duration | Estimated Median Duration |
|---|---|---|---|---|---|---|---|
| Scenario 1 | 98 | 6.8 | 2 | 13.3 | 11 | 14 | 16 |
| Scenario 2 | 99.8 | 13.5 | 0.2 | 14.7 | 14 | 15 | 16 |
| Scenario 3 | 99.3 | 3.7 | 0.7 | 16.0 | 14 | 16 | 18 |
| Scenario 4 | 100 | 84.1 | 0 | 8.0 | 8 | 8 | 8 |
| Scenario 5 | 99.9 | 4.8 | 0.1 | 9.7 | 9 | 10 | 12 |
| *Scenario 6* | *97* | *8.9* | *3* | *11.0* | *10* | *11* | *14* |
| *Scenario 7* | *96* | *42.2* | *4* | *16.4* | *15* | *16* | *18* |
| *Scenario 8* | *95.2* | *14.2* | *4.8* | *15.3* | *8* | *15* | *17* |
| Scenario 9 | 100 | 1.6 | 0 | 12.6 | 13 | 14 | 14 |
| Scenario 10 | 100 | 0.5 | 0 | 15.2 | 16 | 17 | 17 |
| *Scenario 11* | *95.5* | *64* | *4.5* | *17.1* | *17* | *17* | *18* |
| Scenario 12 | 99.6 | 32 | 0.4 | 11.3 | 11 | 12 | 13 |
| Scenario 13 | 97.7 | 22.9 | 2.3 | 8.2 | 8 | 9 | 11 |
| *Scenario 14* | *97.4* | *3.2* | *2.6* | *16.0* | *9* | *15* | *18* |
| Scenario 15 | 99.6 | 4.2 | 0.4 | 12.6 | 12 | 13 | 15 |
| *Scenario 16* | *96.4* | *19.8* | *3.6* | *12.1* | *11* | *12* | *14* |

Table F: Results from using method 2.4, Bootstrap duration CI, targeting an acceptability frontier estimand. Partial and full power, Type-1 Error, Real minimum duration, minimum, 2.5% percentile and median duration recommended. In italics: scenarios for which type 1 error was not strictly controlled within 2.5%.

|  | Partial Power (%) | Full Power (%) | Type 1 Error (%) | True Min Duration | Estimated Minimum Duration | Estimated 2.5th Perc Duration | Estimated Median Duration |
|---|---|---|---|---|---|---|---|
| Scenario 1 | 98.7 | 3 | 1.3 | 14.8 | 12 | 15 | 18 |
| Scenario 2 | 99.8 | 6.3 | 0.2 | 16.0 | 15 | 16 | 18 |
| Scenario 3 | 96.3 | 14.9 | 1.8 | 17.6 | 16 | 18 | 19 |
| Scenario 4 | 100 | 86.1 | 0 | 8.0 | 8 | 8 | 8 |
| Scenario 5 | 99.9 | 2.9 | 0.1 | 9.9 | 9 | 10 | 13 |
| Scenario 6 | 98 | 3.6 | 2 | 11.6 | 10 | 12 | 15 |
| Scenario 7 | 99.3 | 27.9 | 0.7 | 17.8 | 17 | 18 | 19 |
| *Scenario 8* | *92.2* | *28.4* | *7.5* | *17.3* | *8* | *17* | *19* |
| Scenario 9 | 100 | 0.2 | 0 | 12.7 | 13 | 14 | 15 |
| Scenario 10 | 100 | 0 | 0 | 15.4 | 17 | 17 | 18 |
| Scenario 11 | 86.9 | 6.2 | 0.3 | 17.8 | 17 | 18 | 19 |
| Scenario 12 | 99.7 | 20.8 | 0.3 | 11.4 | 11 | 12 | 13 |
| *Scenario 13* | *96.3* | *24.9* | *3.7* | *8.1* | *8* | *8* | *11* |
| *Scenario 14* | *47.1* | *5.2* | *4.4* | *17.8* | *8* | *17* | *19* |
| Scenario 15 | 99.3 | 2.7 | 0.7 | 13.6 | 12 | 14 | 17 |
| Scenario 16 | 98.4 | 9.6 | 1.6 | 12.5 | 11 | 13 | 15 |

Table G: Results from using method 2.4, Bootstrap duration CI, targeting a maximum gradient estimand. Partial and full power, Type-1 Error, Real minimum duration, minimum, 2.5% percentile and median duration recommended. In italics: scenarios for which type 1 error was not strictly controlled within 2.5%. Power is not available for scenario 14, for which the gradient is too steep even at the longest duration.

|  | Partial Power (%) | Full Power (%) | Type 1 Error (%) | True Min Duration | Estimated Minimum Duration | Estimated 2.5th Perc Duration | Estimated Median Duration |
|---|---|---|---|---|---|---|---|
| *Scenario 1* | *67* | *15.3* | *29.8* | *13.939* | *8* | *20* | *15* |
| *Scenario 2* | *62.5* | *30.3* | *36.9* | *16.848* | *12* | *20* | *17* |
| *Scenario 3* | *0* | *42.7* | *57.3* | *19.152* | *8* | *20* | *19* |
| Scenario 4 | 93.2 | 70.6 | 0 | 8 | 8 | 20 | 8 |
| *Scenario 5* | *77.3* | *55* | *19* | *11.152* | *10* | *20* | *12* |
| *Scenario 6* | *90.1* | *32.3* | *4.7* | *11.636* | *8* | *20* | *13* |
| *Scenario 7* | *0* | *42.6* | *57.4* | *19.273* | *8* | *20* | *19* |
| *Scenario 8* | *12* | *42.8* | *57.2* | *18.788* | *8* | *20* | *17* |
| Scenario 9 | 23 | 0.1 | 0 | 13.818 | 14 | 20 | 20 |
| Scenario 10 | 86.6 | 0 | 0 | 16.242 | 18 | 20 | 19 |
| *Scenario 11* | *0* | *63.1* | *36.9* | *19.152* | *18* | *20* | *20* |
| Scenario 12 | 85 | 25.5 | 0.1 | 12.848 | 12 | 20 | 14 |
| *Scenario 13* | *73.3* | *32* | *14* | *9.333* | *8* | *20* | *11* |
| *Scenario 14* | *NA* | *NA* | *52.7* | *20* | *8* | *20* | *18.5* |
| *Scenario 15* | *46* | *23.9* | *50.3* | *14.303* | *8* | *20* | *14* |
| *Scenario 16* | *46* | *36.7* | *27.1* | *14.061* | *13* | *20* | *15* |

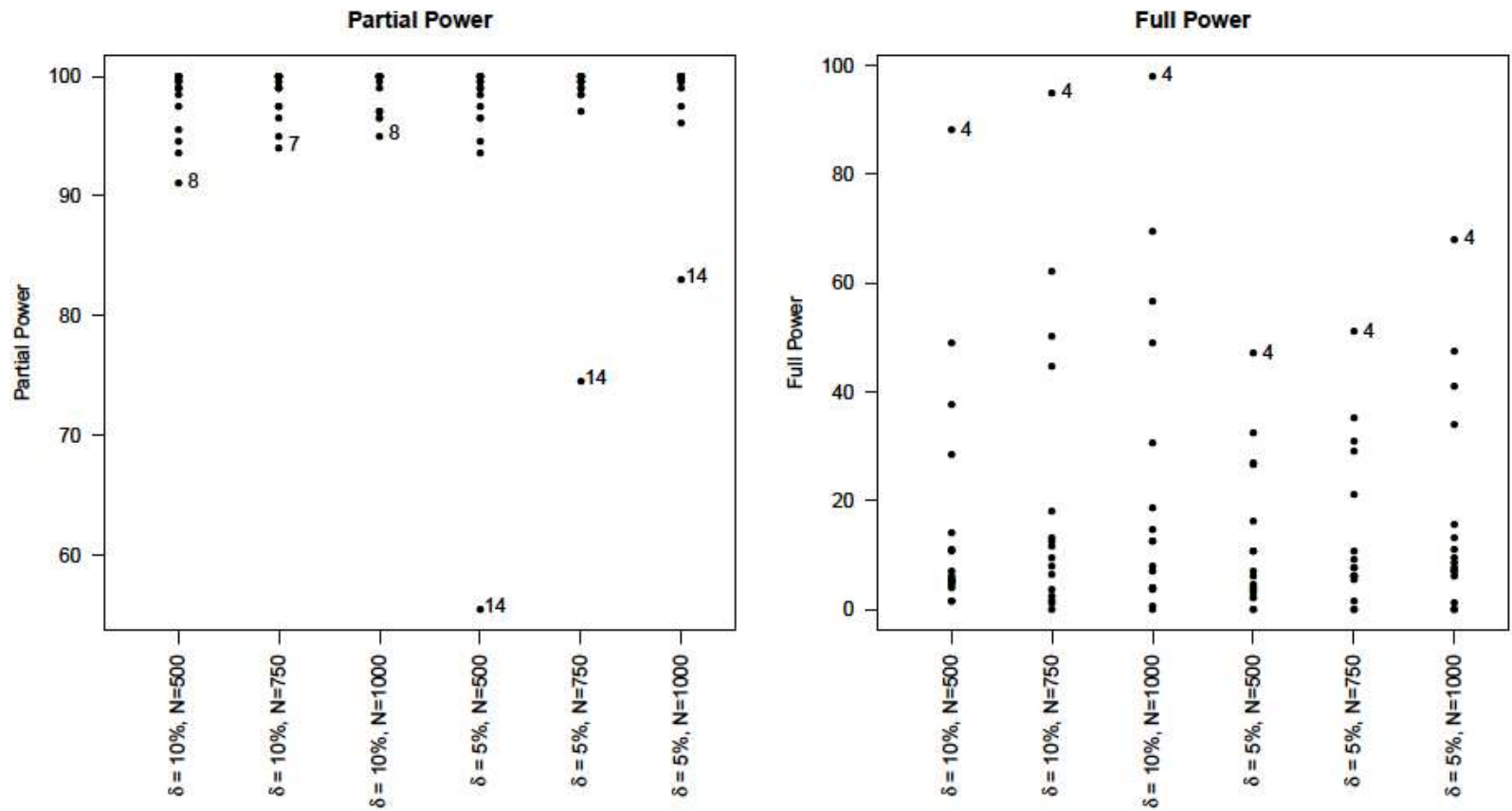

Figure A: Partial and Full Power using bootstrap duration CI (2.4) with 6 different designs. The worst scenario in terms of partial power for each design and the best in terms of full power are indexed.

**Table H: Simulation Scenarios.**

| Type / Equation | Plot | Type / Equation | Plot |
|---|---|---|---|
| 1) Linear Duration-Response curve on log-odds scale: $$\pi = \frac{e^{0.85+0.17(D-8)}}{1+e^{0.85+0.17(D-8)}}$$ | | 9) Logistic Growth Model – Early Growth: $$\pi = 0.05 + \frac{0.9}{1+e^{23-2D}}$$ | |
| 2) Quadratic + linear relation on log-odds scale: $$\pi = \frac{e^{0.62+0.13(D-8)+0.01(D-8)^2}}{1+e^{0.62+0.13(D-8)+0.01(D-8)^2}}$$ | | 10) Logistic Growth Model – Later Growth: $$\pi = 0.05 + \frac{0.9}{1+e^{28-2D}}$$ | |
| 3) Quadratic relation on log-odds scale: $$\pi = \frac{e^{0.85+0.01(D-8)^2}}{1+e^{0.85+0.01(D-8)^2}}$$ | | 11) Gompertz Curve A: $$\pi = 0.9 e^{-e^{-0.5(D-13)}}$$ | |
| 4) Constant response: $$\pi = 0.95$$ | | 12) Gompertz Curve B: $$\pi = 0.9 e^{-e^{-(D-9)}}$$ | |
| 5) Logarithmic relation on log-odds scale: $$\pi = \frac{e^{0.85+1.19\log(D-8)}}{1+e^{0.85+1.19\log(D-8)}}$$ | | 13) Gompertz Curve C: $$\pi = 0.9 e^{-e^{-2(D-7)}}$$ | |
| 6) Square rooted relation on log-odds scale: $$\pi = \frac{e^{0.62+0.67\sqrt{D-8}}}{1+e^{0.62+0.67\sqrt{D-8}}}$$ | | 14) Quadratic on probability scale: $$\pi = 0.7 + 0.01(D-8)^2$$ | |
| 7) Cubic relation on log-odds scale: $$\pi = \frac{e^{1.10+0.002(D-8)^3}}{1+e^{1.10+0.002(D-8)^3}}$$ | | 15) Quadratic on probability scale: $$\pi = 0.7 - 0.01(D-8)^2 + 0.04(D-8)$$ | |
| 8) Cubic + Quadratic relation on log-odds scale: $$\pi = \frac{e^{1.39+0.002(D-8)^2+0.001(D-8)^3}}{1+e^{1.39+0.002(D-8)^2+0.001(D-8)^3}}$$ | | 16) Linear Spline: D<11: $\pi = 0.5 + 0.10(D-8)$ D>11 & D<14: $\pi = 0.8 + 0.04(D-11)$ D>14: $\pi = 0.94 + 0.01(D-14)$ | |

## Appendix A. Delta Method Confidence Interval.

Suppose we ran our fractional polynomial algorithm estimating our duration-response curve:

$$\pi(D) = logit^{-1}(\hat{\alpha} + \hat{\beta}D + \hat{\gamma}D^2),$$

where a linear and a quadratic term have been selected by the algorithm, and $(\hat{\alpha}, \hat{\beta}, \hat{\gamma})$ is the vector of estimated model parameters. Our goal is to obtain a confidence interval for the difference in efficacy between two specific durations $D_1$ and $D_2$. This can be written as a function of model parameters, as:

$$f(\alpha, \beta, \gamma) = \pi(D_1) - \pi(D_2) = logit^{-1}(\alpha + \beta D_1 + \gamma D_1^2) - logit^{-1}(\alpha + \beta D_2 + \gamma D_2^2).$$

Let $\mathbf{\Omega}$ be the variance covariance matrix of the three regression parameters, so that approximately:

$$\begin{pmatrix} \alpha \\ \beta \\ \gamma \end{pmatrix} \sim N\left( \begin{pmatrix} \hat{\alpha} \\ \hat{\beta} \\ \hat{\gamma} \end{pmatrix}, \mathbf{\Omega} \right)$$

And let the Jacobian of function $f(\alpha, \beta, \gamma)$ be the vector of partial derivatives with respect to the three variables:

$$\mathbf{J}_f(\alpha, \beta, \gamma) = \left( \frac{\partial}{\partial \alpha} f, \frac{\partial}{\partial \beta} f, \frac{\partial}{\partial \gamma} f \right) = \left( \frac{e^{\alpha+\beta D_1+\gamma D_1^2}}{(e^{\alpha+\beta D_1+\gamma D_1^2})^2} - \frac{e^{\alpha+\beta D_2+\gamma D_2^2}}{(e^{\alpha+\beta D_2+\gamma D_2^2})^2} \right.,$$

$$\left. D_1 \frac{e^{\alpha+\beta D_1+\gamma D_1^2}}{(e^{\alpha+\beta D_1+\gamma D_1^2})^2} - D_2 \frac{e^{\alpha+\beta D_2+\gamma D_2^2}}{(e^{\alpha+\beta D_2+\gamma D_2^2})^2}, D_1^2 \frac{e^{\alpha+\beta D_1+\gamma D_1^2}}{(e^{\alpha+\beta D_1+\gamma D_1^2})^2} - D_2^2 \frac{e^{\alpha+\beta D_2+\gamma D_2^2}}{(e^{\alpha+\beta D_2+\gamma D_2^2})^2} \right)$$

Then, using the delta method, we have that:

$$f(\alpha, \beta, \gamma) = \pi(D_1) - \pi(D_2) \xrightarrow{d} N\left( f(\hat{\alpha}, \hat{\beta}, \hat{\gamma}), \mathbf{J}_f(\hat{\alpha}, \hat{\beta}, \hat{\gamma}) \mathbf{\Omega} \mathbf{J}_f(\hat{\alpha}, \hat{\beta}, \hat{\gamma})^T \right)$$